\documentclass[12pt,preprint]{aastex}

\usepackage{amssymb}

\begin{document}


\title{Can the Subsonic Accretion Model Explain the Spin Period Distribution of Wind-fed X-ray Pulsars?}


\author{Tao Li, Yong Shao, and Xiang-Dong Li}
\affil{Department of Astronomy, Nanjing University, Nanjing 210023,
China}
\affil{Key Laboratory of Modern Astronomy and Astrophysics (Nanjing University), Ministry of Education, Nanjing 210023, China}
\email{lixd@nju.edu.cn}




\begin{abstract}
Neutron stars in high-mass X-ray binaries (HMXBs) generally accrete from the wind matter of their massive companion stars. Recently Shakura et al. (2012) suggested a subsonic accretion model for low-luminosity ($<4\times 10^{36}$ ergs$^{-1}$), wind-fed X-ray pulsars. To test the feasibility of this model, we investigate the spin period distribution of wind-fed X-ray pulsars with a supergiant companion star, using a population synthesis method. We find that the modeled distribution of supergiant HMXBs in the spin period - orbital period diagram is consistent with observations provided that the winds from the donor stars have relatively low terminal velocities ($\lesssim 1000$ kms$^{-1}$). The measured wind velocities in several supergiant HMXBs seem to favor this viewpoint. The predicted number ratio of wind-fed X-ray pulsars with persistent X-ray luminosities higher and lower than $4\times 10^{36}$ ergs$^{-1}$ is about $1:10$.
\end{abstract}

\keywords{stars: early-type --- stars: evolution --- stars: neutron
--- X-rays: binaries}

\section{Introduction}
High-mass X-ray binaries (HMXBs) contain a neutron star (NS) or a black hole (BH) that is accreting from an O or B type donor star. There are two subtypes of HMXBs, one consisting of a supergiant and the other a B-emission (Be) star \citep[see][for reviews]{vdh2009,reig2011}. The NSs are generally young ($<10^7$ yr) and strongly magnetized, and can be observed as X-ray pulsars due to accretion channeled by their magnetic field lines onto the polar caps, so the X-ray pulsations reflect the NS's spin. Most supergiant HMXBs are close binaries with orbital periods less than a few days,  and the NSs capture the donor's radially-expanding wind matter as outlined by \citet{bon1944}, but in a few of them  Roche-lobe overflow can also occur with an accretion disk formed around the NSs. The Be systems are usually in wide, eccentric orbits, and the NSs accrete from the Be star's disklike winds, usually near periastron.

\citet{cor1984,cor1985,cor1986} noticed that in the spin period ($P_{\rm s}$) vs. orbital period ($P_{\rm orb}$) diagram the distribution of various subtypes of HMXBs shows distinct features: wind-fed NSs in supergiant HMXBs have much longer spin periods than disk-fed NSs, while the Be systems seem to roughly follow a positive correlation between $P_{\rm s}$ and $P_{\rm orb}$. Moreover, both $P_{\rm s}$ and $P_{\rm orb}$ in the Be systems seem to show a bimodal distribution \citep{knigge2011}. These
features must be related to the properties of the wind and the accretion processes in HMXBs, and have been extensively investigated \citep[e.g.,][]{ste1986,heu1987,wat1989,kin1991,li1996,zhang2004,dai2006,che2014}.

The classical picture of the NS spin evolution in supergiant HMXBs was constructed by \citet{dav1981}, who argued that the captured wind matter forms a quasi-static, spherical atmosphere around the NS. Generally speaking, the spin-down evolution of a NS before steady accretion occurs contains at least two phases: the radio pulsar (or ejector) phase and the propeller phase. A newborn NS is usually rapidly rotating and emits energetic particles. If the accreting material is stopped outside the light cylinder, the NS acts as a radio pulsar and its spin-down is caused by energy loss via magnetic dipole radiation. When the accreting matter can permeate the NS's light cylinder, the propeller phase begins and the NS experiences rapid spin-down due to the interaction between the NS magnetosphere and the surrounding plasma \citep{pri1972}. Although the efficiency of the propeller spin-down torque is still in debate \citep[e.g.,][]{ill1975,dav1979,wan1985,mor2003}, it is generally thought that an equilibrium spin period will be reached at the end of the propeller phase, when the net torque exerted on the NS vanishes \citep{Bhattacharya1991}.

In the subsequent quasi-spherical accretion phase, the freely-falling matter is decelerated by a shock formed above the NS magnetosphere. The condition of steady accretion is determined by whether the magnetosphere boundary is unstable with respect to Rayleigh-Taylor instability \citep{arons1976} or Kelvin-Helmholtz instability \citep{burnard1983}. \citet{sha2012} argued that, if the X-ray luminosity $L_{\rm X} > 4 \times 10^{36}\,{\rm erg\,s^{-1}}$, the shocked matter can efficiently cool via Compton processes and enter
the magnetosphere; if $L_{\rm X} < 4 \times 10^{36}\,{\rm erg\,s^{-1}}$, a subsonic settling accretion regime sets in, with a hot convective shell formed above the NS magnetosphere. In the latter case, both the accretion torque and the equilibrium spin period  deviate from those in the traditional spherical wind accretion model. This subsonic accretion model has been applied to account for the observational characteristics of some long-spin period X-ray pulsars including \object{GX 1+4} \citep{gon2012}, \object{4U 1954+31} \citep{mar2011}, \object{SXP 1062} \citep{pop2012}, \object{X Per} \citep{lut2012}, and  GX 301$-$2 \citep{postnov2015}, as well as supergiant fast X-ray transients (SFXTs) \citep{sha2014}, which are a subclass of HMXBs  characterized by sporadic X-ray flares with peak luminosities $\sim 10^{36}-10^{37}$ ergs$^{-1}$ \citep{duc2014}. However, a study on the global feature of HMXBs in this model is still lacking.

This paper aims to test the model of \citet{sha2012} by
comparing the expected spin period distribution based on population synthesis calculation with the observations of wind-fed HMXBs. Different from \citet{dai2006,dai2016}, we consider the NS spin evolution both prior to and during the HMXB phase, in which the subsonic accretion may dominate. In Section 2 we describe our assumptions and theoretical considerations. Section 3 presents the calculated results, which are then compared with observations.  Our discussion and conclusions are in Section 4.

\section{Model}
\subsection{Formation of Incipient HMXBs}

The formation history of HMXBs starts from primordial massive binaries and has been  reviewed by \citet{Bhattacharya1991} and \citet{vdh2009}. The more massive, primary star first evolves, fills its Roche-lobe, transfers matter to the secondary, and finally leaves a NS (or a BH) with a supernova (SN) explosion. If the binary is not disrupted after the SN, the NS will gravitationally capture the wind matter from its companion star and become an X-ray source.
We use the rapid binary evolution code developed by \citet{hur2000,hur2002} and updated by \citet{kie2006} to generate the incipient HMXB population. This code combines the standard assumptions for primordial binaries with analytic prescriptions that describe stellar evolution, binary interactions and SN explosions. \citet{shao2014} further modified the code for the formation and evolution of NSs and BHs. We briefly describe them as follows.

If the primary star is significantly more massive than the secondary, the mass transfer may be dynamically unstable, leading to common envelope (CE) evolution, which is still not well understand. To judge the stability of mass transfer, we adopt the calculated results by \citet{shao2014} for a grid of binaries rather the empirical relations in \citet{hur2002} for the critical mass ratios, and assume a $50\%$ mass transfer efficiency \citep[see][for details]{shao2014}. We use the energy conservation formalism  for CE evolution  \citep{web1984}, by equating the difference in the orbital energy with the binding energy of the primary's envelope,
\begin{equation}
\alpha_{\rm CE}\left(\frac{GM_{\rm 1,f}M_2}{2a_{\rm f}}-\frac{GM_{\rm 1,i}M_2}{2a_{\rm i}}\right)
=\frac{GM_{\rm 1,i}M_{\rm 1,e}}{\lambda R_{\rm L1}},
\end{equation}
where $\alpha_{\rm CE}\leq 1$ is an efficiency parameter, $G$ is the gravity constant, $M_1$, $M_{\rm 1,e}$, $M_{\rm 1,f}$ are the total mass, the envelope mass, and the core mass of the primary, respectively,  $M_2$ is the secondary mass, $a$ is the binary separation, $R_{\rm L1}$ is the Roche-lobe radius of the primary at the onset of Roche-lobe overflow, the indices i and f denote the values at the beginning and end of the CE stage, respectively, and the parameter $\lambda$ reflects the effect of the mass distribution within the envelope and the contribution from other energies besides gravitational energy \citep[e.g.,][]{koo1990,dew2000,pod2003}. We used the calculated values of $\lambda$  by \citet{xu2010} for high- and intermediate-mass stars at various evolutionary stages, and take $\alpha_{\rm CE}=1.0$ in our calculation.

We consider NSs born from both core-collapse and electron-capture SNe \citep{pod2004,vdh2004} which are associated with the collapse of an ONeMg core of intermediate-mass stars \citep{nomoto1984}, and adopt the mass range for the progenitors of  electron-capture SNe according to \citet{bel2008} and \citet{fry2012}. We assume that the newborn NSs receive a kick because of asymmetric SN explosions and its direction is isotropically distributed. For the magnitude of the kick velocity we adopt a Maxwellian distribution with one-dimensional rms velocity $\sigma_{\rm k}=265\, {\rm kms^{-1}}$ for core-collapse SNe \citep{hob2005} and $\sigma_{\rm k}=50\, {\rm kms^{-1}}$  for electron-capture SNe \citep{des2006}.

The initial parameters in our binary population synthesis calculation are taken as follows. All the binaries are tidally circularized \citep{hur2002}. The primary mass $M_1$  is in the range of $7-60 M_\odot$, following the \citet{kro1993} mass function. The secondary mass $M_2$ lies between $3M_\odot$ and $30M_\odot$, with the initial mass ratio $M_2/M_1$ uniformly distributed between 0 and 1. The logarithm of the orbital separation $\ln a$ follows a uniform distribution, with the values of $a$ ranging from $3R_\odot$ to $10^4R_\odot$. We adopt Solar metallicity for the stars, and assume that the star formation proceeds at a constant rate ($5M_\odot {\rm yr^{-1}}$) during the past 13 Gyr.

\begin{figure}
\plotone{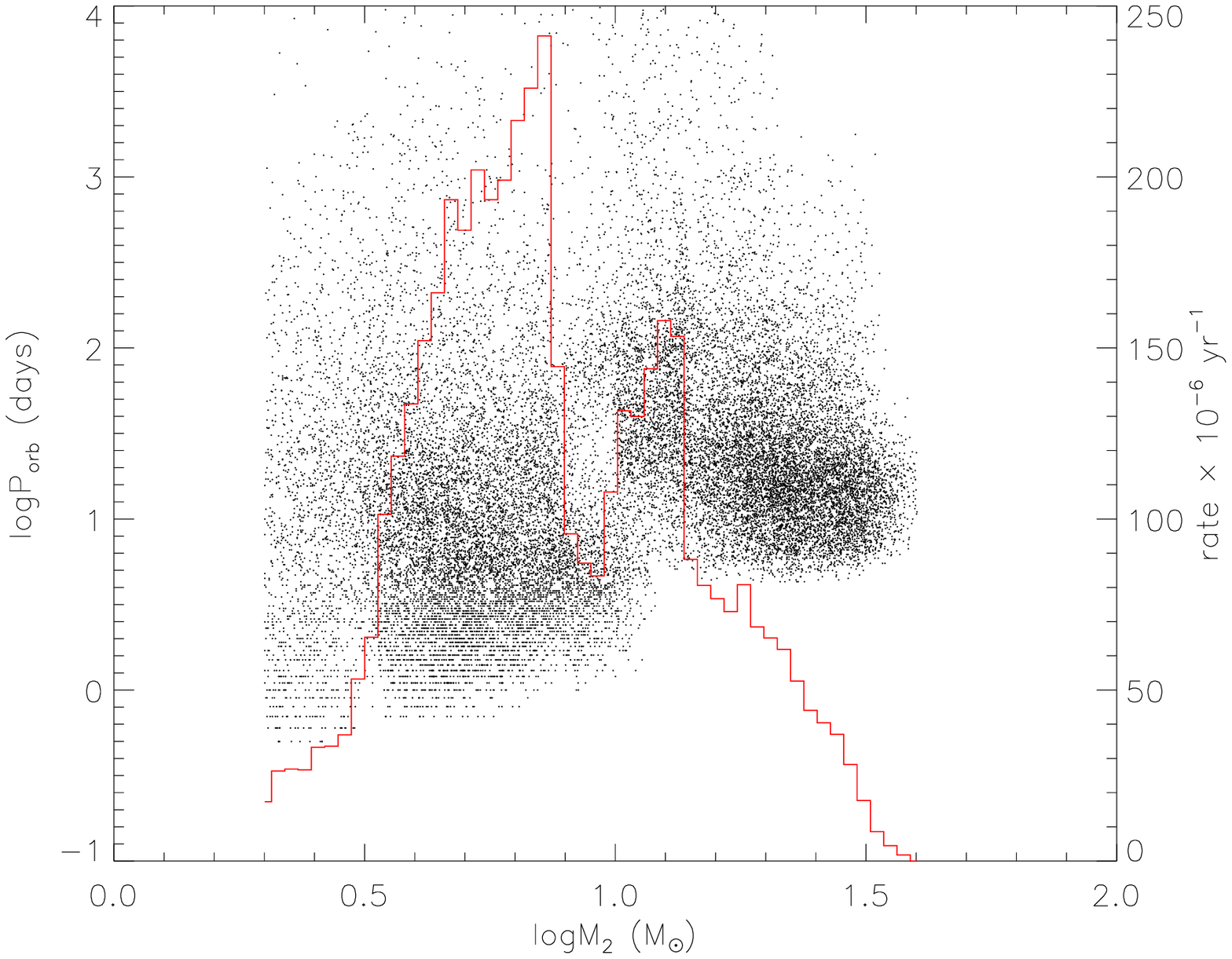}
\caption{Distributions of incipient HMXBs in the initial companion mass ($M_2$) vs. the orbital period ($P_{\rm orb}$) plane. The red curve represents their birthrate as a function of $M_2$. \label{fig1}}
\end{figure}

Figure \ref{fig1} demonstrates the distribution of $P_{\rm orb}$ and $M_2$ for the produced incipient NS binaries. To compare with the observed HMXBs we select the $4\times 10^5$ binaries with $P_{\rm orb}<1000$ days and $10M_{\odot}\leq M_2\leq 30M_{\odot}$ for further investigation. We also calculate the luminosities, radii and effective temperatures of the companion stars, and use them to estimate the wind mass loss rates and the mass transfer rates.

\subsection{Spin Evolution of NSs}
We then investigate the spin history of a $1.4\,M_\odot$ NS with an OB companion star based on the model of \citet{dav1981} but with considerable modifications. As briefly described below, the NS evolves through three different evolutionary stages: the radio pulsar, propeller, and accretor phases.

Normally a NS is born with a short spin period and a strong magnetic field. Its magnetic or radiation pressure is able to
to expel the wind matter outside the Bondi accretion radius $r_{\rm G}=2GM/v^2$ \citep{bon1952}, or the light cylinder radius, $r_{\rm lc}=cP_{\rm s}/2\pi$. Here $M$ is the NS mass and $v=(v_{\rm orb}^2+v_{\rm w}^2)^{1/2}$ is the velocity of the wind relative to the NS, where $v_{\rm orb}$ and $v_{\rm w}$ are the orbital velocity of the NS and the wind velocity at the NS's orbit, respectively. The spin-down torque induced by magnetic dipole radiation in this radio pulsar phase is
\begin{equation}
I\dot{\Omega}_{\rm s}=-\frac{2}{3}\frac{\mu^2\Omega_s^3}{c^3},
\end{equation}
where $I$, $\mu$, and  $\Omega_{\rm s}$ are the moment of inertia,  the magnetic dipole moment, and  the angular velocity of the NS, respectively. The radio pulsar phase ends when the wind matter enters the Bondi radius $r_{\rm G}$ or the light cylinder radius $r_{\rm lc}$  \citep[see][for details]{dav1981}, and the spin periods are correspondingly,
\begin{equation}
P_{\rm a} \simeq 0.8 \mu_{30}^{1/3} \dot{M}_{15}^{-1/6}
(M/M_{\odot})^{1/3}v_8^{-5/6}\,{\rm s},
\end{equation}
\begin{equation}
P_{\rm b}\simeq
1.2\dot{M}_{15}^{-1/4}\mu_{30}^{1/2}v_8^{-1/2}\,{\rm s},
\end{equation}
where $\mu_{30}=\mu/10^{30}\,{\rm Gcm^3}$, $v_8=v/10^8\,{\rm cm\,s^{-1}}$, and $\dot{M}=10^{15}\dot{M}_{15}$ \,g\,s$^{-1}$ is the mass transfer rate given by \citep{bon1944}
\begin{equation}
\label{dm}
\dot{M}=\pi r_{\rm G}^2\rho_{\rm w}v,
\end{equation}
where $\rho_{\rm w}$ is the density of the companion's wind matter at the NS's orbit. For an isotropically-expanding wind we have
\begin{equation}
\label{rho}
\rho_{\rm w}=-\dot{M}_2/(4\pi a^2v_{\rm w}).
\end{equation}
Here $\dot{M}_2$ is the mass loss rate of the companion star, which can be estimated with the empirical formula proposed by \citet{nie1990} for OB type stars,
\begin{equation}
-\dot{M}_2=9.6\times10^{-15}
(R_2/R_\odot)^{0.81}(L_2/L_\odot)^{1.24}(M_2/M_\odot)^{0.16}\,M_{\odot}{\rm yr}^{-1},
\label{ml}
\end{equation}
where $L_2$ and $R_2$ are the luminosity and the radius of the companion star, respectively.

In the propeller phase the infalling matter starts to interact with the magnetosphere, but the fast rotating magnetosphere inhibits the matter from steady accretion onto the NS.  At this moment the magnetospheric radius $r_{\rm m}=[\mu^4/(2GM\dot{M}^2)]^{1/7}$ \citep{pri1972} is still
larger than the corotation radius $r_{\rm co}=(GM/\Omega_{\rm
s}^2)^{1/3}$. The infalling matter is stopped and then expelled at the magnetosphere because of the centrifugal barrier, extracting the angular momentum of the NS. The efficiency of angular momentum loss due to
the propeller mechanism is still in debate \citep{ill1975,wan1985,ikh2007,rom2005,ust2006,dan2010}. When the ejected mater is corotating with the magnetosphere, the spin-down torque is
\begin{equation}
I\dot{\Omega}_{\rm s}=-\dot{M}r_{\rm m}^2\Omega_{\rm s}.
\end{equation}
This phase ends when $r_{\rm co}=r_{\rm m}$, and the corresponding spin period
\begin{equation}
\label{peq}
P_{\rm eq}\simeq
23\mu_{30}^{6/7}\dot{M}_{15}^{-3/7}(M/M_{\odot})^{-5/7}\,{\rm s}
\end{equation}
is called the equilibrium period.

In the following accretor phase the NS spin period will be also changed due to mass and angular momentum transfer. However, both observations \citep{bil1997} and numerical simulations \citep[e.g.,][]{mat1987,fry1988,saw1989,ruf1994,bp2009,don2011,cruz2012} indicate erratic spin-up/down with relatively low averaged angular momentum transfer rate during wind accretion. Thus, one may expect that the spin periods of the NSs during the accretor phase do not change much from the period reached in the propeller phase. Therefore, \citet{dai2006} stopped their calculation when the companion star evolves off the main-sequence, no matter whether $P_{\rm eq}$ is reached. This simple picture should be modified, because, according to \citet{sha2012}, there exist two cases of quasi-spherical accretion depending on the accretion rate. The matter heated up in the shock above the magnetosphere can fall toward the NS surface only when it cools down rapidly, which requires the X-ray luminosity $\gtrsim 4\times10^{36}\,{\rm erg\,s^{-1}}$. In this regime, the NS spin evolution is determined by the magnitude of the angular momentum carried by the captured matter as discussed above.
If the X-ray luminosity $< 4\times10^{36}\,{\rm erg\,s^{-1}}$, the shocked matter is unable to cool down, a quasi-static shell is formed around the NS magnetosphere, in which the hot matter settles down subsonically. In this subsonic accretion regime the accretion rate is determined by the ability of the matter to enter the magnetosphere via instabilities, which can be considerably lower than the capture rate by the NS. The torques exerted on the NS come from both accretion and the magnetosphere-plasma interaction. In the case of moderate magnetosphere-plasma coupling, the equation that governs the NS spin evolution reads
\begin{equation}
I\dot{\Omega}_{\rm s} = A\dot M^{7/11} - B\dot M^{3/11},
\end{equation}
where the coefficients $A$ and $B$ are (in CGS units)
\begin{equation}
A \simeq 4.60\times 10^{31}K_1\mu_{30}^{1/11}v_8^{-4}\left(\frac{P_{\rm
orb}}{10\,\hbox{d}}\right)^{-1},
\end{equation}
and
\begin{equation}
B \simeq 5.49\times
10^{32}K_1\mu_{30}^{13/11}\left(\frac{P_{\rm
s}}{100\,\hbox{s}}\right)^{-1}
.
\end{equation}
Here $K_1$ is a dimensionless numerical factor with typical value of $40$  \citep{sha2012}. The equilibrium spin period  in the subsonic accretion phase is
\begin{equation}
\label{pseq}
P_{\rm s,eq} \simeq 1193\mu_{30}^{12/11}v_8^{4}\dot M_{16}^{-4/11}\left(\frac{P_{\rm orb}}{10\,\hbox{d}}\right)\,{\rm s}.
\end{equation}
Obviously, NSs experiencing subsonic accretion can have much longer spin period than those undergoing rapid accretion at the same rate.

\section{Results}
We then calculate how the NS spins change in the incipient HMXBs, based on the theoretical model presented in Section 2. We assume that the  NSs  initially spin at a period uniformly distributed between $10$ ms and $100$ ms. A log-normal distribution is set for the initial NS magnetic fields (in units of Gauss), with a mean of 12.5 and a standard deviation of 0.3. We assume that the fields do not decay during the lifetime of a HMXB. We adopt the standard \citet{cas1975} formula for the wind velocity $v_{\rm w}$,
\begin{equation}
v_{\rm w}=v_\infty(1-R_2/a)^\beta,
\end{equation}
where $v_\infty$ is the terminal velocity of the wind, and $\beta \sim 0.5-1$, taken to be 0.8 in this work. The mass capture rate depends on many parameters, among which the relative velocity $v$ and hence the wind velocity $v_{\rm w}$ are the most sensitive ones. So we set $v_\infty = v_{\rm esc}$ and $3v_{\rm esc}$ to examine its influence \citep[e.g.,][]{wat1989,owo2014}, where $v_{\rm esc}$ is the escape velocity at the surface of the companion star.

Since our calculations are limited to the stage of quasi-spherical wind accretion, we keep the binary evolution going on until the companion star starts to fill its Roche lobe. Our calculation shows that the produced NS populations are distributed in the radio pulsar, propeller, and accretor phases, depending on the activity of the NS magnetic field - wind interaction. To compare with observations of wind-fed NSs in HMXBs, we chose the calculated results for NSs that have entered the accretor phase with X-ray luminosities $\geq 10^{33}\,{\rm erg\,s^{-1}}$, which are regarded to be potential X-ray pulsars.  The accretion processes in Be/X-ray binaries are quite different from supergiant systems, because of the complicated structure of the Be star's wind and eccentric orbits, which imply that the quasi-spherical accretion model may not apply for these sources \citep[see][for detailed discussion]{dai2006,che2014}. So we only pay attention to the wind-fed X-ray pulsars in the supergiant systems.

Figure~\ref{fig2} displays the calculated spin period distribution, in which the left and right panels correspond to the cases of $v_\infty = v_{\rm esc}$ and $3v_{\rm esc}$, respectively. The darkness of each element represents the relative number of the X-ray binaries. The diamonds mark the observed wind-fed supergiant HMXBs with known $P_{\rm s}$ and $P_{\rm orb}$, and the triangles represent disk-fed X-ray pulsars  \citep[data from][and references therein]{tow2011,dai2016}. According to our calculation, the wind-fed X-ray pulsars are clustered into the upper and lower subgroups, related to different regimes of quasi-spherical accretion. Since the NS mass capture rates are higher for shorter orbital periods, only HMXBs in relatively narrow orbits can have X-ray luminosities $\gtrsim 4 \times 10^{36}\,{\rm erg\,s^{-1}}$. The NSs in these systems are experiencing rapid accretion, but their spin periods still stay around the periods reached during the propeller phase, because they have got small net angular momentum during the accretor phase. So they are located at the lower-left shaded region of the diagram. The other sources are undergoing the subsonic accretion and have longer spin periods, located in the upper part of the diagram.

We find that when $v_\infty = v_{\rm esc}$, around $8\%$ of the NSs are in the rapid accretion stage, and $89\%$ and $3\%$ in the subsonic accretion stage with X-ray luminosities $>$ and $< 10^{33}\,{\rm erg\,s^{-1}}$, respectively (the latter are not displayed in the diagram). If $v_\infty$ is increased to $3v_{\rm esc}$, the mass transfer rates are lowered by a factor of $\sim 80$ according to Eqs.~(\ref{dm}) and (\ref{rho}). The luminosity condition for the subsonic accretion becomes easier to satisfy, further extending the subsonic accretion region in the diagram, and the corresponding numbers of the NSs become $\sim 1\%$, $40\%$ and $59\%$, respectively. As indicated by Eq.~(\ref{pseq}), the equilibrium period for subsonic accretion is very sensitive to the relative wind velocity. The faster the wind, the longer the periods. This makes the spin periods of NSs in the subsonic accretion region in the right panel are nearly 2 orders of magnitude longer than in the left panel. Obviously, the results of $v_\infty = v_{\rm esc}$ better match the observations.

To demonstrate the difference in the expected HMXB distribution in different models, we recalculate the NS spin evolution but without the subsonic accretion process taken into account, and show the results in Fig.~\ref{fig3}.  This was previously done by \citet{dai2006}, but they did not consider the NSs that enter the accretor phase when the companion stars have evolved to be supergiants.
The left and right panel also correspond to $v_\infty = v_{\rm esc}$ and $3v_{\rm esc}$, respectively. Since the equilibrium period in the rapid accretion regime is much shorter than in the subsonic accretion regime at the same $v_\infty$, most of the shaded regions in Fig.~\ref{fig3} are lower than in Fig.~\ref{fig2}, except the ones with X-ray luminosities $> 4 \times 10^{36}\,{\rm erg\,s^{-1}}$, which are located at the same positions.  It seems that the results in the right panel with $v_\infty = 3v_{\rm esc}$ better fit the observations. In this case, the percentages of the sources with X-ray luminosities $L_{\rm X}>4 \times 10^{36}\,{\rm ergs^{-1}}$ and $10^{33}\,{\rm ergs^{-1}}<L_{\rm X}< 4 \times 10^{36}\,{\rm ergs^{-1}}$ are $\sim 1\%$ and $52\%$ respectively, while the rest $47\%$ have X-ray luminosities $L_{\rm X}< 10^{33}\,{\rm ergs^{-1}}$.

\section{Discussion and conclusions}
We have applied the model proposed by \citet{sha2012} for quasi-spherical accretion to explore the spin period distribution for X-ray pulsars in supergiant HMXBs. One of the differences between this model and the traditional one is that, some of the NSs may experience long-term spin-down during the subsonic accretion phase, and have much longer equilibrium period. So to explain the current spin periods, it is not required that the spin-down processes were conducted when the NS was interacting with relatively weak winds from a main-sequence companion star \citep{wat1989}. The two different equilibrium periods (i.e., Eqs.~[13] and [9]) imply that, given the same $P_{\rm orb}$, the NSs in different accretion regimes can have distinct spin period distributions.

\begin{figure}
\plottwo{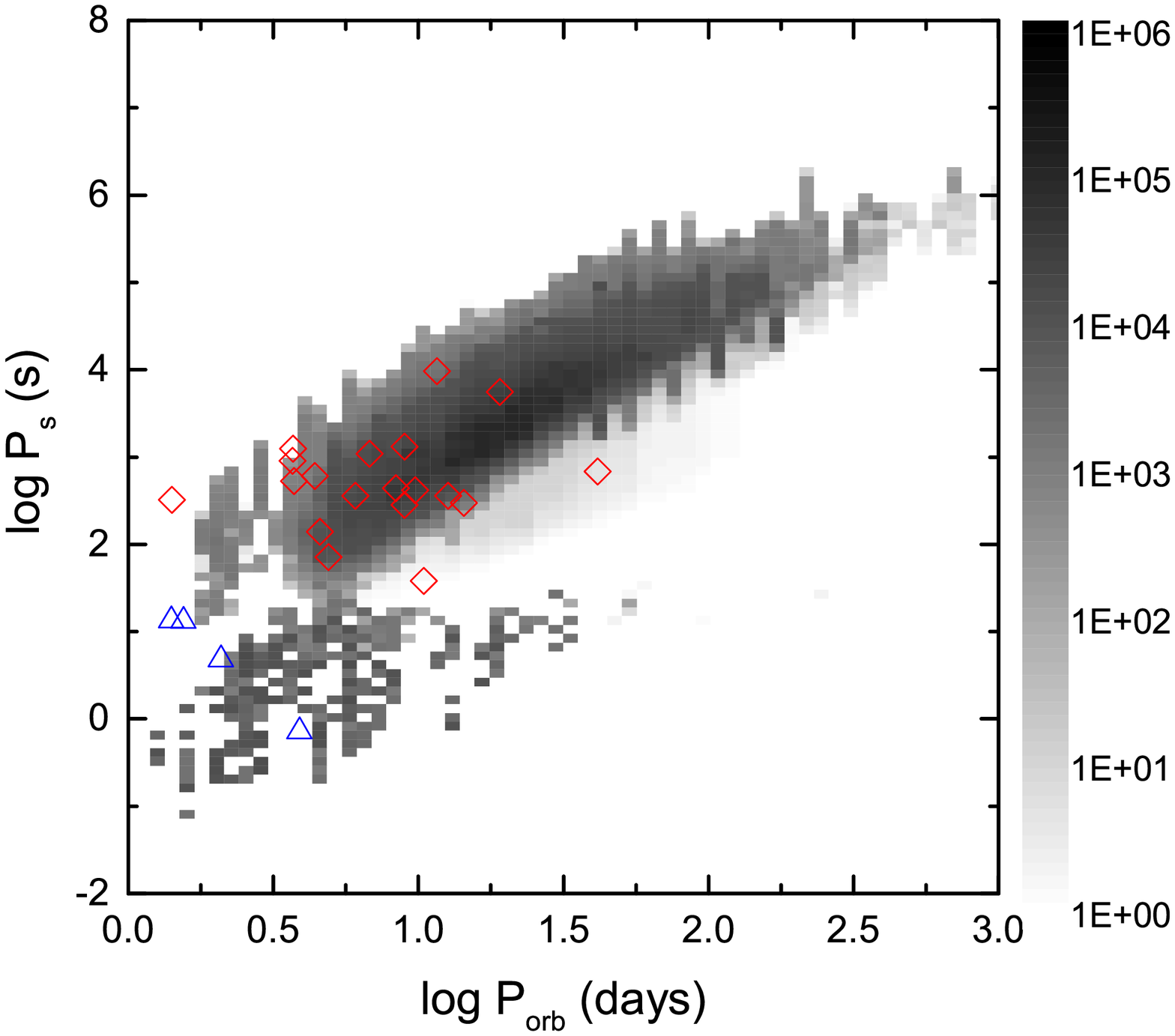}{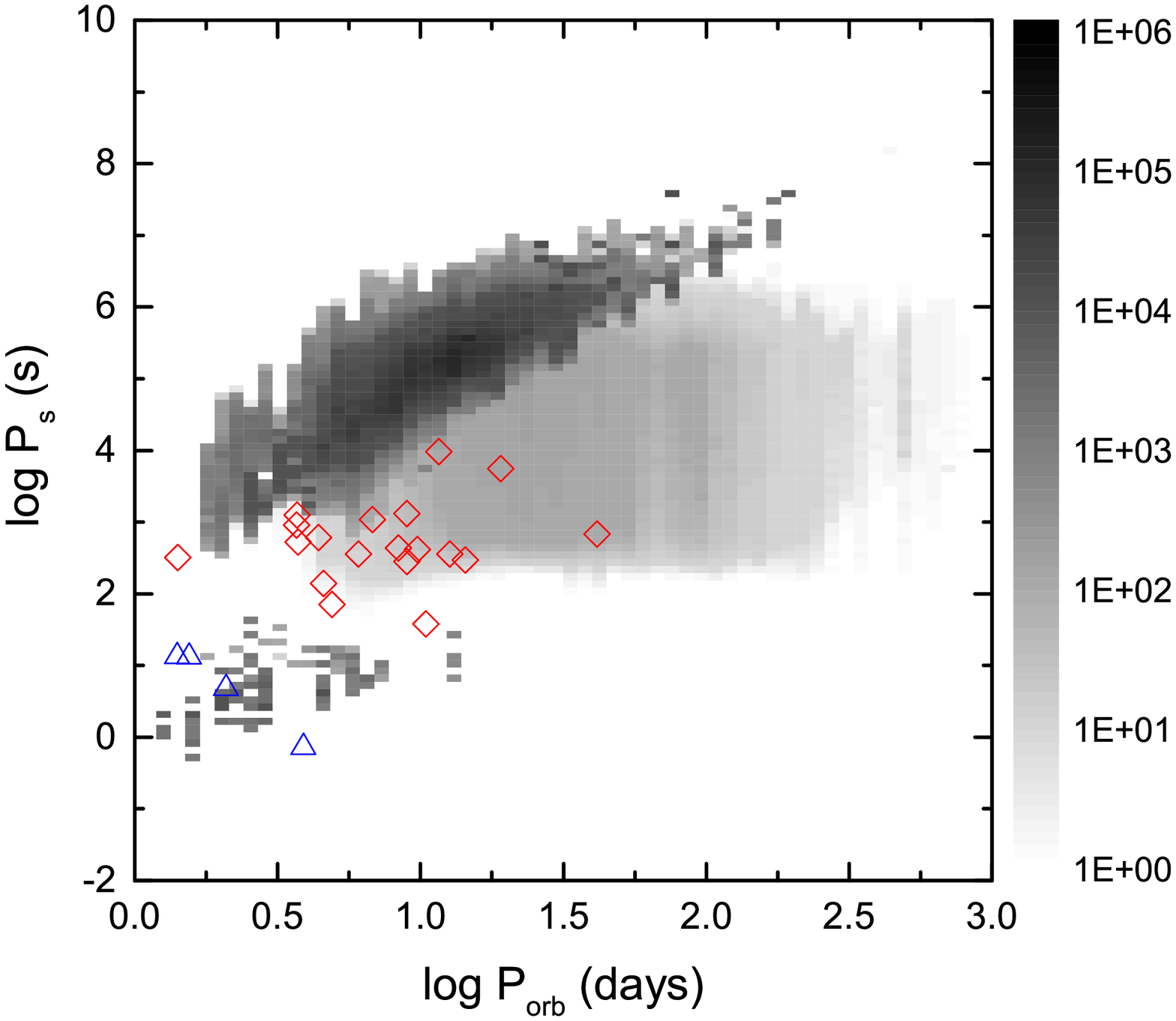}
\caption{The distribution of wind-fed HMXBs in the $P_{\rm s}-P_{\rm orb}$ diagram, with the subsonic settling accretion taken into account. The gray scale denotes the relative number of HMXBs. The diamonds and triangles represent the observed wind-fed and Roche-lobe overflowing supergiant HMXBs, respectively. The left and right panels correspond to $v_\infty = v_{\rm esc}$ and $3v_{\rm esc}$, respectively. \label{fig2}}
\end{figure}

\begin{figure}
\plottwo{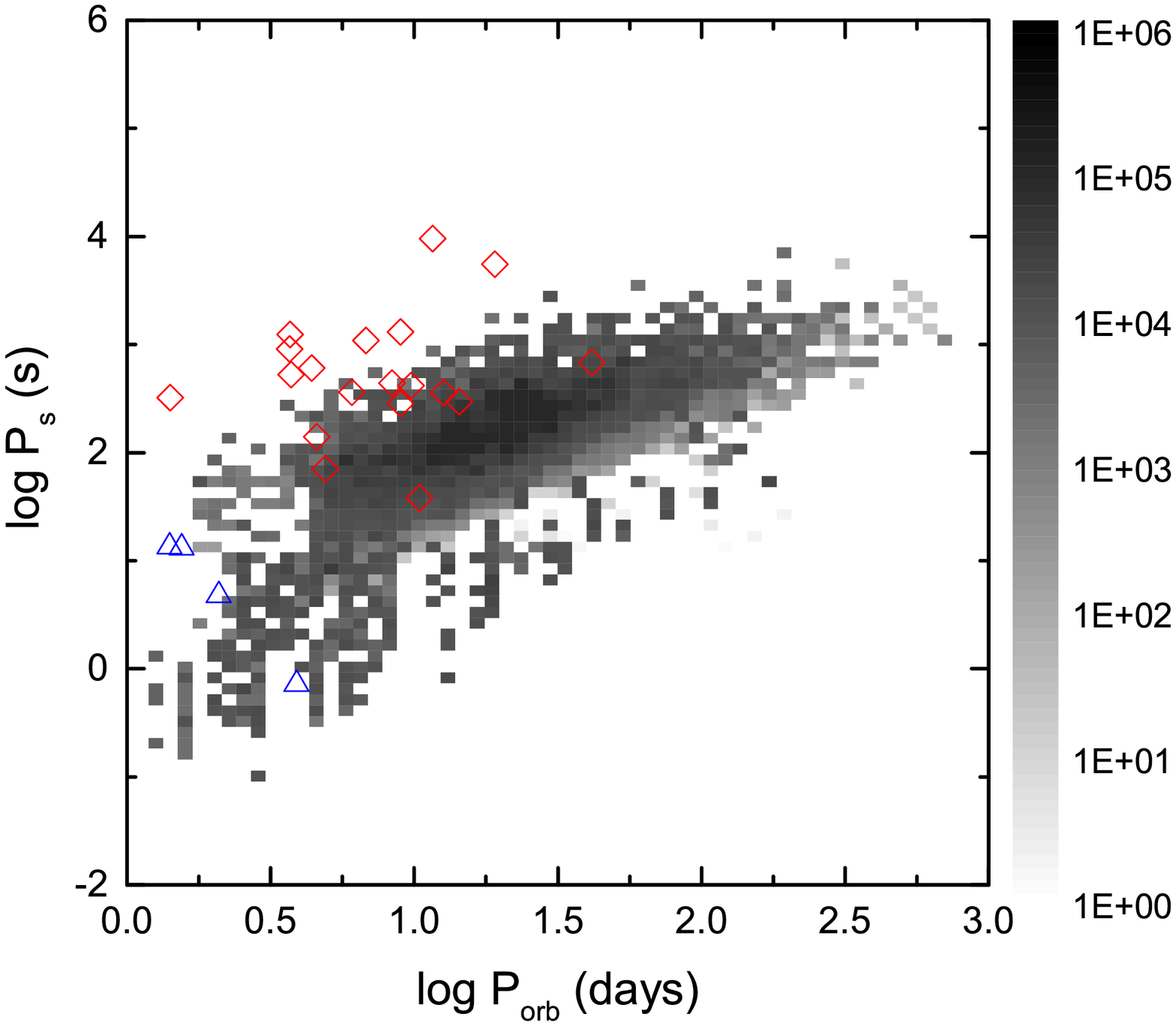}{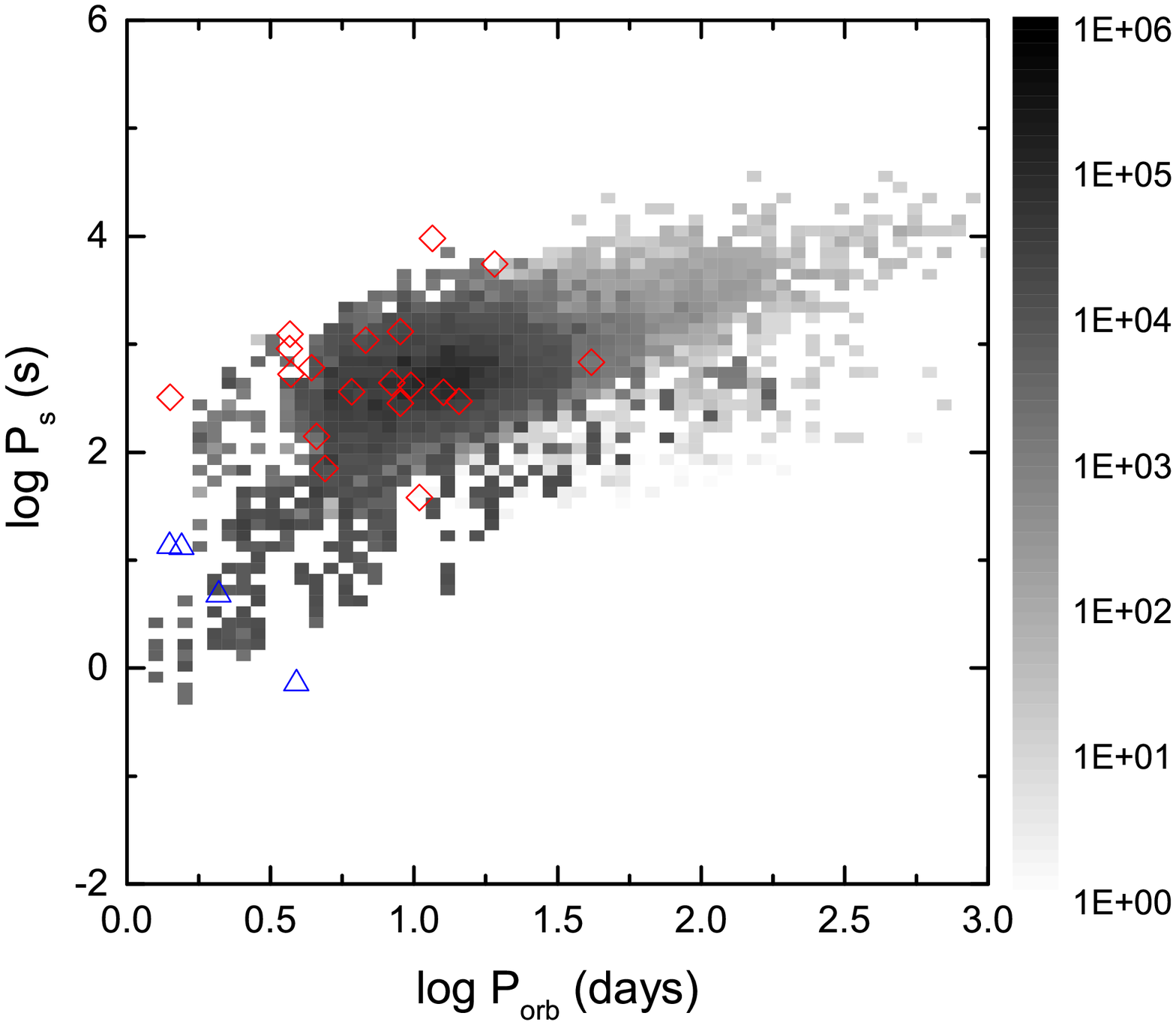}
\caption{Same as Fig. \ref{fig2} but without the subsonic accretion taken into account. \label{fig3}}
\end{figure}

Whether the subsonic accretion is taken into account or not, our calculated results displayed in both Figs.~\ref{fig2} and \ref{fig3} can be in accord with the observed supergiant HMXBs in the $P_{\rm s}-P_{\rm orb}$ diagram. The difference lies in that, to match the observations, the model with subsonic accretion requires a relative low wind velocity $v_\infty \sim v_{\rm esc}$ (with the average value $\sim 500-800\,{\rm km\,s^{-1}}$), otherwise a higher wind velocity $v_\infty \sim 3v_{\rm esc}$ (with the average value $\sim 1500-2400\,{\rm km\,s^{-1}}$) is required. Since the mass capture rates strongly depend on the wind velocity, the values of $v_\infty$ can serve as a plausible indicator to discriminate the models. They have been estimated by modelling of the ultraviolet resonance lines, and have been extensively investigated  \citep[e.g.,][]{how1989,pri1990,lam1995,how1997,pri1998,pul1996,kud1999}.
Generally $v_\infty/v_{\rm esc}\simeq 2.6$ for Galactic single supergiants of types earlier than B1 \citep{kud2000}. However, this ratio  was found to be considerably lower (down to $\lesssim 1$) in the OB supergiants
in HMXBs \citep{loon2001}. The most reliable values of $v_\infty$ for HMXBs which we have found in the literature are as follows: $\sim 600-700$ kms$^{-1}$ for Vela X-1, $\sim 1700$ kms$^{-1}$ for 4U 1700$-$37, $\sim 500$ kms$^{-1}$ for LMC X-4, $\lesssim 600$ kms$^{-1}$ for SMC X$-$1 \citep{loon2001}, $\sim 1700$ kms$^{-1}$ for 4U 1907$+$09 \citep{cox2005}, $1500\pm 200$ kms$^{-1}$ for IGR J17544$-$2619 \citep{gg2016}, $\sim 350$ kms$^{-1}$ for 4U2206$+$54 \citep{ribo2006}, $\sim 305$ kms$^{-1}$ for GX 301$-$2 \citep{kaper2006}, and $500\pm 100$ kms$^{-1}$ for 4U 1908$+$075 \citep{mar2015}. Thus quite a few systems have $v_\infty\sim v_{\rm esc}$, lending support to the subsonic accretion model.

Our results also indicate that NSs expected to experience subsonic accretion are overwhelming in the whole population, which can also be tested by observations.  According to our calculated results, the number ratio of supergiant HMXBs with $L_{\rm X}$ above and below $4\times 10^{36}$ ergs$^{-1}$ is $1:11$ in the subsonic accretion model with $v_\infty = v_{\rm esc}$, and $1:52$ in the traditional model with $v_\infty = 3v_{\rm esc}$. However, almost all HMXBs are variable in X-rays with a dynamical range from a few to $10^5$ \citep{ste1986,reig2011}. Long-term monitoring the sampled HMXBs and reasonable evaluation of the X-ray luminosities will also help resolve the issue.

In our calculations we employ the classical $\beta$-velocity law to model the supergiant's wind. Modern self-consistent calculations of the wind hydrodynamics in Wolf-Rayed stars show that the wind acceleration is more complicated: the wind structure shows two acceleration regions, one close to the hydrostatic wind base in the optically thick part of the atmosphere, and the other farther out in the wind \citep{gra2005,gra2008}. Thus a simple $\beta$-velocity law may not be an accurate description of the velocity field, and the strong acceleration in the inner part is best described by $\beta\sim 1$ whereas the outer part tends towards higher $\beta$.
This type of velocity profile in the Wolf-Rayet winds has also been found in recent calculations of supergiant winds \citep{sander2015}.
In this case the average wind velocity is a bit lower than that in our calculation with the same terminal velocity $v_\infty$. This will make the shaded regions in Figs.~\ref{fig2} and \ref{fig3} downward slightly, but it does not significant affect our final results.

Another issue is that, because of the kick velocity received by the NSs at their formation,  one does not expect the NS's and/or the supergiant's spins to be aligned with the orbit \citep[e.g.,][]{lai1996,heu1997}. In principle this may not cause considerable changes in the mass and angular momentum transfer rates in the binaries, since the wind material is assumed to be spherically expanding from the donor star and there is not a preferred direction for the wind distribution. Actually most supergiant HMXBs are in close, circular orbits, suggesting strong tidal interactions between the components. The SN kick may play a more important role in Be/X-ray binaries. A misaligned Be star disk is less likely to be tidally truncated by the NS than in coplanar systems, which is more likely to trigger giant outbursts rather than quasi-periodic, normal outbursts \citep{mar2011,oka2013}. This might be related to the bimodal distribution of the spin periods in Be/X-ray binaries \citep{knigge2011,che2014}.

We finally caution that our approximation of a smooth stellar wind should be considered unrealistic. Stellar winds are likely to have both large-scale (quasi)-cyclical structures  \citep{kaper1998} and small-scale, stochastic structures \citep{ro2002,puls2008}, which may be caused by various instabilities in the stars. So our results should be considered as averaged over long-time for the NS - wind interaction.

\acknowledgments

We are grateful to the referee for helpful comments. This work was funded by the Natural Science Foundation of China under grant numbers 11133001 and 11333004, and the Strategic Priority Research Program of CAS (under grant number XDB09000000).

\clearpage

\begin{thebibliography}{}
\bibitem[Arons \& Lea(1976)]{arons1976}Arons J., \& Lea S. M., 1976, \apj, 207, 914
\bibitem[Belczynski et al.(2008)]{bel2008} Belczynski, K., Kalogera, V., Rasio, F.~A., et al.\ 2008, \apjs, 174, 223
\bibitem[{{Bhattacharya \& van den Heuvel} (1991)}]{Bhattacharya1991}
Bhattacharya, D., \& van den Heuvel E. P. J. 1991, Phys. Rep., 203, 1
\bibitem[Bildsten et al.(1997)]{bil1997} Bildsten, L., Chakrabarty, D., Chiu, J., et al.\ 1997, \apjs, 113, 367
\bibitem[Blondin \& Pope(2009)]{bp2009}
Blondin, J. M., \& Pope, T. C. 2009, \apj, 700, 95
\bibitem[Bondi(1952)]{bon1952}
Bondi, H. 1952, \mnras, 112, 195
\bibitem[Bondi \& Hoyle(1944)]{bon1944} Bondi, H., \& Hoyle, F.\ 1944, \mnras, 104, 273
\bibitem[Burnard et al.(1983)]{burnard1983}Burnard D. J., Lea S. M., \& Arons J., 1983, \apj, 266, 175
\bibitem[Castor et al.(1975)]{cas1975} Castor, J.~I., Abbott, D.~C., \& Klein, R.~I.\ 1975, \apj, 195, 157
\bibitem[Cheng et al.(2014)]{che2014} Cheng, Z.-Q., Shao, Y., \& Li, X.-D.\ 2014, \apj, 786, 128
\bibitem[Corbet(1984)]{cor1984} Corbet, R.~H.~D.\ 1984, \aap, 141, 91
\bibitem[Corbet(1985)]{cor1985} Corbet, R.\ 1985, \ssr, 40, 409
\bibitem[Corbet(1986)]{cor1986} Corbet, R.~H.~D.\ 1986, \mnras, 220, 1047
\bibitem[Cox et al.(2005)]{cox2005}
Cox, N. L. J., Kaper, L., Foing, B. H., \& Ehrenfreund, P. 2005, \aap, 436, 661
\bibitem[Cruz-Osorio et al.(2012)]{cruz2012}
Cruz-Osorio, A., Lora-Clavijo, F. D., \& Guzm\'an, F. S. 2012, \mnras, 426, 732
\bibitem[Dai et al.(2006)]{dai2006} Dai, H.-L., Liu, X.-W.,
\& Li, X.-D.\ 2006, \apj, 653, 1410
\bibitem[Dai et al.(2016)]{dai2016} Dai, H.-L., Liu, X.-W., \& Li, X.-D.\ 2016, MNRAS, 457, 3889
\bibitem[D'Angelo \& Spruit(2010)]{dan2010}
D'Angelo, C. R., \& Spruit, H. C. 2010, \mnras, 406, 1208
\bibitem[Davies et al.(1979)]{dav1979} Davies, R.~E., Fabian, A. C., \& Pringle, J.~E.\ 1979, \mnras, 186, 779
\bibitem[Davies \& Pringle(1981)]{dav1981} Davies, R.~E., \& Pringle, J.~E.\ 1981, \mnras, 196, 209
\bibitem[de Kool(1990)]{koo1990} de Kool, M.\ 1990, \apj, 358,
189
\bibitem[Dessart et al.(2006)]{des2006} Dessart, L., Burrows, A., Ott, C.~D., et al.\ 2006, \apj, 644, 1063
\bibitem[Dewi \& Tauris(2000)]{dew2000} Dewi, J.~D.~M., \& Tauris, T.~M.\ 2000, \aap, 360, 1043
\bibitem[D\"onmez et al.(2011)]{don2011}
D\"onmez, O., Zanotti, O., \& Rezzolla, L. 2011, \mnras, 412, 1659
\bibitem[Ducci et al.(2014)]{duc2014}
Ducci, L., Doroshenko, V., Romano, P., Santangelo, A., \& Sasaki, M. 2014,
\aap, 568, 76
\bibitem[Fryer et al.(2012)]{fry2012} Fryer, C.~L., Belczynski,
K., Wiktorowicz, G., et al.\ 2012, \apj, 749, 91
\bibitem[Fryxell \& Taam(1988)]{fry1988} Fryxell, B.~A., \& Taam, R.~E.\ 1988, \apj, 335, 862
\bibitem[Gimenez-Garcia et al.(2016)]{gg2016}
Gimenez-Garcia, A., Shenar, T., Torrejon, J. M., Oskinova, L., \& Martinez-Nunez, S. et al. 2016, \aap, in press (arXiv:1603.00925)
\bibitem[Gonz{\'a}lez-Gal{\'a}n et al.(2012)]{gon2012} Gonz{\'a}lez-Gal{\'a}n, A., Kuulkers, E., Kretschmar, P., et al.\ 2012, \aap, 537, A66
\bibitem[Gr{\"a}fener \& Hamann(2005)]{gra2005} Gr{\"a}fener, G., \& Hamann, W.-R.\ 2005, \aap, 432, 633
\bibitem[Gr{\"a}fener \& Hamann(2008)]{gra2008} Gr{\"a}fener, G., \& Hamann, W.-R.\ 2008, \aap, 482, 945
\bibitem[Hobbs et al.(2005)]{hob2005} Hobbs, G., Lorimer,
D.~R., Lyne, A.~G., \& Kramer, M.\ 2005, \mnras, 360, 974
\bibitem[Howarth \& Prinja(1989)]{how1989} Howarth, I.~D., \& Prinja, R.~K.\ 1989, \apjs, 69, 527
\bibitem[Howarth et al.(1997)]{how1997} Howarth, I.~D.,
Siebert, K.~W., Hussain, G.~A.~J., \& Prinja, R.~K.\ 1997, \mnras, 284, 265
\bibitem[Hurley et al.(2000)]{hur2000} Hurley, J.~R., Pols,
O.~R., \& Tout, C.~A.\ 2000, \mnras, 315, 543
\bibitem[Hurley et al.(2002)]{hur2002} Hurley, J.~R., Tout,
C.~A., \& Pols, O.~R.\ 2002, \mnras, 329, 897
\bibitem[Ikhsanov(2001)]{ikh2001} Ikhsanov, N.~R.\ 2001, \aap, 368, L5
\bibitem[Ikhsanov(2007)]{ikh2007} Ikhsanov, N.~R.\ 2007, \mnras, 375, 698
\bibitem[Illarionov \& Sunyaev(1975)]{ill1975} Illarionov, A.~F., \& Sunyaev, R.~A.\ 1975, \aap, 39, 185
\bibitem[Jiang \& Li(2005)]{jia2005} Jiang, Z.-B., \& Li, X.-D.\ 2005, \cjaa, 5, 487
\bibitem[Kaper et al.(2006)]{kaper2006}
Kaper, L., van der Meer, A., \& Najarro, F., 2006, \aap, 457, 595
\bibitem[Kaper \&  Fullerton(1998)]{kaper1998}
Kaper, L., \&  Fullerton, A. W. 1998, Cyclical variability in stellar winds (Berlin, New York: Springer-Verlag)
\bibitem[Kiel \& Hurley(2006)]{kie2006} Kiel, P.~D., \& Hurley, J.~R.\ 2006, \mnras, 369, 1152
\bibitem[King(1991)]{kin1991} King, A.~R.\ 1991, \mnras, 250,
3
\bibitem[Knigge et al.(2011)]{knigge2011}
Knigge, C., Coe, M. J., \& Podsiadlowski, Ph. 2011, \nat, 479, 372
\bibitem[Kroupa et al.(1993)]{kro1993} Kroupa, P., Tout, C.~A.,
\& Gilmore, G.\ 1993, \mnras, 262, 545
\bibitem[Kudritzki et al.(1999)]{kud1999} Kudritzki, R.~P., Puls, J., Lennon, D.~J., et al.\ 1999, \aap, 350, 970
\bibitem[Lai et al.(1996)]{lai1996} Lai, D., Bildsten, L., \& Kapsi, V. M., 1995, ApJ, 452, 819
\bibitem[Kudritzki \& Puls(2000)]{kud2000} Kudritzki, R.-P., \& Puls, J.\ 2000, \araa, 38, 613
\bibitem[Lamers et al.(1995)]{lam1995} Lamers, H.~J.~G.~L.~M.,
Snow, T.~P., \& Lindholm, D.~M.\ 1995, \apj, 455, 269
\bibitem[Li \& van den Heuvel(1996)]{li1996} Li, X.-D., \& van den Heuvel, E.~P.~J.\ 1996, \aap, 314, L13
\bibitem[Lutovinov et al.(2012)]{lut2012} Lutovinov, A.,
Tsygankov, S., \& Chernyakova, M.\ 2012, \mnras, 423, 1978
\bibitem[Marcu et al.(2011)]{mar2011} Marcu, D.~M., F{\"u}rst, F., Pottschmidt, K., et al.\ 2011, \apjl, 742, L11
\bibitem[Martinz et al.(2011)]{mar2011} Martin, R. G., Pringle, J. E., Tout, C. A., \& Lubow, S. H. 2011, \mnras, 416, 2827
\bibitem[Mart\'inez-N\'u\~nez et al.(2015)]{mar2015}
Mart\'inez-N\'u\~nez, S., Sander, A., G\'imenez-Garc\'ia, A., G\'onzalez-Gal\'an, A., \& Torrej\'on, J. M. 2015, \aap, 578, A107
\bibitem[Matsuda et al.(1987)]{mat1987} Matsuda, T., Inoue, M., \& Sawada, K. 1987, \mnras, 226, 785
\bibitem[Mori \& Ruderman(2003)]{mor2003} Mori, K., \& Ruderman, M.~A.\ 2003, \apjl, 592, L75
\bibitem[Nieuwenhuijzen \& de Jager(1990)]{nie1990} Nieuwenhuijzen, H., \& de Jager, C.\ 1990, \aap, 231, 134
\bibitem[Nomoto(1984)]{nomoto1984} Nomoto, K. 1984, \apj, 277, 791
\bibitem[Okazaki et al.(2013)]{oka2013}
Okazaki, A. T., Hayasaki, K., \& Moritani, Y. 2013, PASJ, 65, 41
\bibitem[Owocki(2014)]{owo2014} Owocki, S.\ 2014, arXiv:1409.2084
\bibitem[Podsiadlowski et al.(2004)]{pod2004} Podsiadlowski, P., Langer, N., Poelarends, A.~J.~T., et al.\ 2004, \apj, 612, 1044
\bibitem[Podsiadlowski et al.(2003)]{pod2003} Podsiadlowski,
P., Rappaport, S., \& Han, Z.\ 2003, \mnras, 341, 385
\bibitem[Popov \& Turolla(2012)]{pop2012} Popov, S.~B., \& Turolla, R.\ 2012, \mnras, 421, L127
\bibitem[Postnov et al.(2015)]{postnov2015}
Postnov, K. A., Mironov, A. I., Lutovinov, A. A., Shakura, N. I., Kochetkova, A. Yu., \& Tsygankov, S. S. 2015, \mnras, 446, 1013
\bibitem[Pringle \& Rees(1972)]{pri1972} Pringle, J.~E., \& Rees, M.~J.\ 1972, \aap, 21, 1
\bibitem[Prinja et al.(1990)]{pri1990} Prinja, R.~K., Barlow,
M.~J., \& Howarth, I.~D.\ 1990, \apj, 361, 607
\bibitem[Prinja
\& Crowther(1998)]{pri1998} Prinja, R.~K., \& Crowther, P.~A.\ 1998, \mnras, 300, 828
\bibitem[Puls et al.(1996)]{pul1996} Puls, J., Kudritzki, R.-P., Herrero, A., et al.\ 1996, \aap, 305, 171
\bibitem[Puls et al.(2008)]{puls2008}
Puls, J., Markova, N., \& Scuderi, S. 2008, in Mass Loss from Stars and the Evolution of Stellar Clusters, eds. A. de Koter, L. J. Smith, \& L. B. F. M. Waters (San Francisco: Astronomical Society of the Pacific),  p. 101
\bibitem[Reig(2011)]{reig2011}Reig, P. 2011, \apss, 322, 1
\bibitem[Rib\'o et al.(2006)]{ribo2006}
Rib\'o, M., Negueruela, I., Blay, P., Torrej\'on, J. M., \&  Reig, P. 2006, \aap, 449, 687
\bibitem[Romanova et al.(2005)]{rom2005}
Romanova, M.M., Ustyugova, G. V. Koleoba, A. V., \& Lovelace, R. V. E., 2005, \apjl, 635, L165
\bibitem[Ruffert(1994)]{ruf1994} Ruffert, M.\ 1994, \apj, 427, 342
\bibitem[Runacres \&  Owocki(2002)]{ro2002}
Runacres, M. C., \&  Owocki, S. P. 2002, \aap, 381, 1015
\bibitem[Sander et al.(2015)]{sander2015}Sander, A., Shenar, T., Hainich, R., G\'imenez-Garc\'ia, A., Todt, H., \& Hamann, W.-R. 2015, \aap, 577, A13
\bibitem[Sawada et al.(1989)]{saw1989}
Sawada, K., Matsuda, T., Anzer, U., Borner, G. E., \& Livio, M. 1989, \aap, 231,
263
\bibitem[Shakura et al.(2012)]{sha2012} Shakura, N., Postnov,
K., Kochetkova, A., \& Hjalmarsdotter, L.\ 2012, \mnras, 420, 216
\bibitem[Shakura et al.(2014)]{sha2014} Shakura, N., Postnov,
K., Sidoli, L., \& Paizis, A. \ 2014, \mnras, 442, 2325
\bibitem[Shao \& Li(2014)]{shao2014} Shao, Y., \& Li, X.-D.\ 2014, \apj, 796, 37
\bibitem[Stella et al.(1986)]{ste1986} Stella, L., White,
N.~E., \& Rosner, R.\ 1986, \apj, 308, 669
\bibitem[Tauris \& van den Heuvel(2006)]{tau2006} Tauris, T.~M., \& van den Heuvel, E.~P.~J.\ 2006, Compact stellar X-ray sources, 623
\bibitem[Townsend et al.(2011)]{tow2011} Townsend, L.~J., Coe, M.~J., Corbet, R.~H.~D., \& Hill, A.~B.\ 2011, \mnras, 416, 1556
\bibitem[Ustyugova et al.(2006)]{ust2006}
Ustyugova, G. V., Koleoba, A. V., Romanova, M.M., \& Lovelace, R. V. E. 2006, \apj, 646, 304
\bibitem[van den Heuvel(2004)]{vdh2004} van den Heuvel, E. P. J. 2004, in ESA Special Publication, Vol. 552, 5th INTEGRAL Workshop on the INTEGRAL Universe, ed. V. Schoenfelder, G. Lichti, \& C. Winkler (Noordwijk: ESA), 185
\bibitem[van den Heuvel(2009)]{vdh2009} van den Heuvel, E.~P.~J. 2009, in Physics of Relativistic Objects in Compact Binaries: From Birth to Coalescence, Astrophysics and Space Science Library, vol. 359, eds.  M. Colpi, P. Casella, V. Gorini, U. Moschella,  \&  A. Possenti (Springer Netherlands), p. 125
\bibitem[van den Heuvel \& Rappaport(1987)]{heu1987} van den Heuvel, E.~P.~J., \& Rappaport, S.\ 1987, IAU Colloq.~92: Physics of Be Stars, 291
\bibitem[van den Heuvel \& van Paradijs(1997)]{heu1997} van den Heuvel, E. P. J., \& van Paradijs, J. 1997, \apj, 483, 399
\bibitem[van Loon et al.(2001)]{loon2001}
van Loon, J. Th., Kaper, L., \& Hammerschlag-Hensberge, G. 2001, \aap, 375, 498
\bibitem[Wang \& Robertson(1985)]{wan1985} Wang, Y.-M., \& Robertson, J.~A.\ 1985, \aap, 151, 361
\bibitem[Waters \& van Kerkwijk(1989)]{wat1989} Waters, L.~B.~F.~M., \& van Kerkwijk, M.~H.\ 1989, \aap, 223, 196
\bibitem[Webbink(1984)]{web1984} Webbink, R.~F.\ 1984, \apj,
277, 355
\bibitem[Xu \& Li(2010)]{xu2010} Xu, X.-J., \& Li, X.-D.\ 2010, \apj, 716, 114
\bibitem[Zhang et al.(2004)]{zhang2004} Zhang, F., Li, X.-D., \& Wang, Z.-R.\ 2010, ChJA\&A, 4, 320

\end{thebibliography}
\end{document}